\documentclass[namedreferences]{solarphysics}
\usepackage[hyperref,optionalrh,showbiblabels]{spr-sola-addons}
\usepackage{graphicx}
\usepackage{color}           
\usepackage{breakurl}        
\usepackage{multirow}

\begin{document}
\begin{article}
\begin{opening}

\title{A new inner heliosphere proton parameter data set from the Helios mission}

\author[addressref={aff1},corref,email={david.stansby14@imperial.ac.uk}]{\inits{D.}\fnm{David}~\lnm{Stansby}}
\author[addressref={aff2}]{\inits{C.}\fnm{Chadi}~\lnm{Salem}}
\author[addressref={aff3}]{\inits{L.}\fnm{Lorenzo}~\lnm{Matteini}}
\author[addressref={aff1}]{\inits{T. S.}\fnm{Timothy}~\lnm{Horbury}}

\address[id=aff1]{Department of Physics, Imperial College London, London, SW7 2AZ, United Kingdom}
\address[id=aff2]{Space Sciences Laboratory, University of California, Berkeley, CA 94720, USA}
\address[id=aff3]{LESIA, Observatoire de Paris, Universit\'e PSL, CNRS, Sorbonne Universit\'e, Univ. Paris Diderot, Sorbonne Paris Cit\'e, 5 place Jules Janssen, 92195 Meudon, France}

\begin{abstract}
In the near future, Parker Solar Probe and Solar Orbiter will provide the first comprehensive in-situ measurements of the solar wind in the inner heliosphere since the Helios mission in the 1970s. We describe a reprocessing of the original Helios ion distribution functions to provide reliable and reproducible data to characterise the proton core population of the solar wind in the inner heliosphere. A systematic fitting of bi-Maxwellian distribution functions was performed to the raw Helios ion distribution function data to extract the proton core number density, velocity, and temperatures parallel and perpendicular to the magnetic field.  We present radial trends of these derived proton parameters, forming a benchmark from which new measurements in the inner heliosphere will be compared to. The new dataset has been made openly available for other researchers to use, along with the source code used to generate it.
\end{abstract}

\end{opening}

\section{Introduction}
With the imminent launches of Parker Solar Probe \citep{Fox2015} and Solar Orbiter \citep{Muller2013}, heliospheric and solar physics are about to enter a new age of discovery. To date, the most comprehensive mission to visit the inner heliosphere and make in-situ measurements of the solar wind was the Helios mission, consisting of two spacecraft, which explored the heliosphere from 0.3 AU - 1 AU in the 1970s and 1980s, covering solar minimum between solar cycles 20 and 21 and the maximum of solar cycle 21 \citep{PORSCHE1977}. The data returned by these spacecraft provided a wealth of information, however the computational resources available to process the data were limited at the time, and there is no publicly available dataset containing reliable and reproducible moments derived from the full 3D distribution functions. In this paper we revisit the plasma measurements made on board the two Helios spacecraft. The plasma data have been reprocessed before \cite[e.g.][]{Marsch2004, Matteini2007, Hellinger2011}, but importantly the new data set and the code used to generate it is openly available to researchers. This makes the dataset easily reproducible and reusable.

The solar wind primarily consists of protons, with a smaller fraction of alpha particles ($\sim$ 1\% - 5\%), a series of other minor ions ($\ll$ 1\%), and neutralising electrons \citep{Neugebauer1962, Marsch1982c, Pilipp1987, Bochsler2007}. The proton population can be further split into two: the proton core which accounts for $\sim$ 90\% of the protons, and the smaller proton beam which travels at a different velocity to the core \citep{Feldman1973, Marsch1982b}. Here we present systematic bi-Maxwellian fits the proton core population for the entire duration the Helios mission. 

In section \ref{sec:previous data} we give a brief overview of the data that was already widely available to researchers. In section \ref{sec:raw data} an overview of the plasma instrumentation is given, and in section \ref{sec:fitting} an overview of the data processing is given and the new dataset summarised. In section \ref{sec:comparison} we compare the dataset to the previously available data. In section \ref{sec:trends} we use the new dataset to provide a new set of radial trends for the proton core population of the solar wind.

\section{Previously available data}
\label{sec:previous data}
As far as we know, the only other publicly available set of proton plasma parameters available from the Helios mission is the ``merged" data set\footnote{Available at \url{ftp://cdaweb.gsfc.nasa.gov/pub/data/helios/helios1/merged/} and \url{ftp://cdaweb.gsfc.nasa.gov/pub/data/helios/helios2/merged/}}. This set of parameters were calculated in the 1970s and 1980s by taking numerical moments of 1D energy spectra, obtained by integrating the 3D distributions over all solid angles. Although taking numerical moments is computationally fast, it has a number of restrictions:
\begin{itemize}
	\item The total number density is calculated; this does not discriminate between the proton core and beam populations.
	\item Only the component of the temperature tensor in the radial direction ($T_{r}$) is calculated. For a bi-Maxwellian with two true temperatures, $T_{r}$ depends in a non-trivial way on both temperatures, but also the angle the instantaneous magnetic field vector makes with the radial direction.
\end{itemize}
In the rest of this paper we use \texttt{moment} to refer to the previously available data, and \texttt{corefit} to refer to the reprocessed dataset described here. We stress that our method of reprocessing is not intrinsically better than taking moments, but instead provides a different set of information describing the properties of individual distribution functions, which complements the information already available.

\section{Data processing}
\label{sec:processing}
\subsection{Raw data}
\label{sec:raw data}
Both Helios 1 and 2 were equipped with an experiment for measuring the distribution function of positively charged particles in the solar wind, called the E1 plasma instrument \citep{Schwenn1975}. For much more detailed information we refer the reader to the instrument technical paper \citep{Rosenbauer2018}.

The E1 experiment was an electrostatic analyser that counted particles as a function of their energy per charge ($E/q$). There were 32 $E/q$ channels logarithmically spaced between 0.155 kV and 15.3 kV, and 9 angular elevation channels oriented perpendicular to the spin plane of the spacecraft (the ecliptic plane) and separated by 5$^{\circ}$. Resolution in the azimuthal direction was built up using the spin of the spacecraft, with measurements taken every 5$^{\circ}$. During each spin period the flux in each angular bin was measured at a fixed $E/q$. Over 32 spins of the spacecraft this allowed all 32 $E/q$ channels to be sampled in each angular direction. In high resolution mode a 7 x 7 grid of angular measurements in all 32 $E/q$ channels was transmitted back to Earth, centred around the distribution peak; in low resolution mode this was reduced to a 5 x 5 angular grid across 9 $E/q$ channels, again centred on the distribution peak. Distributions transmitted in both modes contain enough data for locating and fitting to the proton core.

Because the E1 instrument had no mass discrimination, the 3D distribution functions contain contributions from both protons and alpha particles \citep{Marsch1982c}. Because the protons and alphas are well separated in energy, and the protons form the majority of the distribution, it was simple to fit a bi-Maxwellian distribution to just the protons.

Both spacecraft also had two magnetometers: the E2 experiment with data available at 4 vectors/second \citep{Musmann1975} and the E3 experiment with data available at one vector every 6 seconds \citep{Scearce1975}. Magnetic field data was used as part of the fitting process to constrain the symmetry axis of the fitted bi-Maxwellian. For times when the higher rate E2 data was available it was used, but otherwise data from the E3 experiment was used.

\subsection{Fitting process}
\label{sec:fitting}

\begin{figure}
	\centerline{\includegraphics[width=0.8\textwidth]{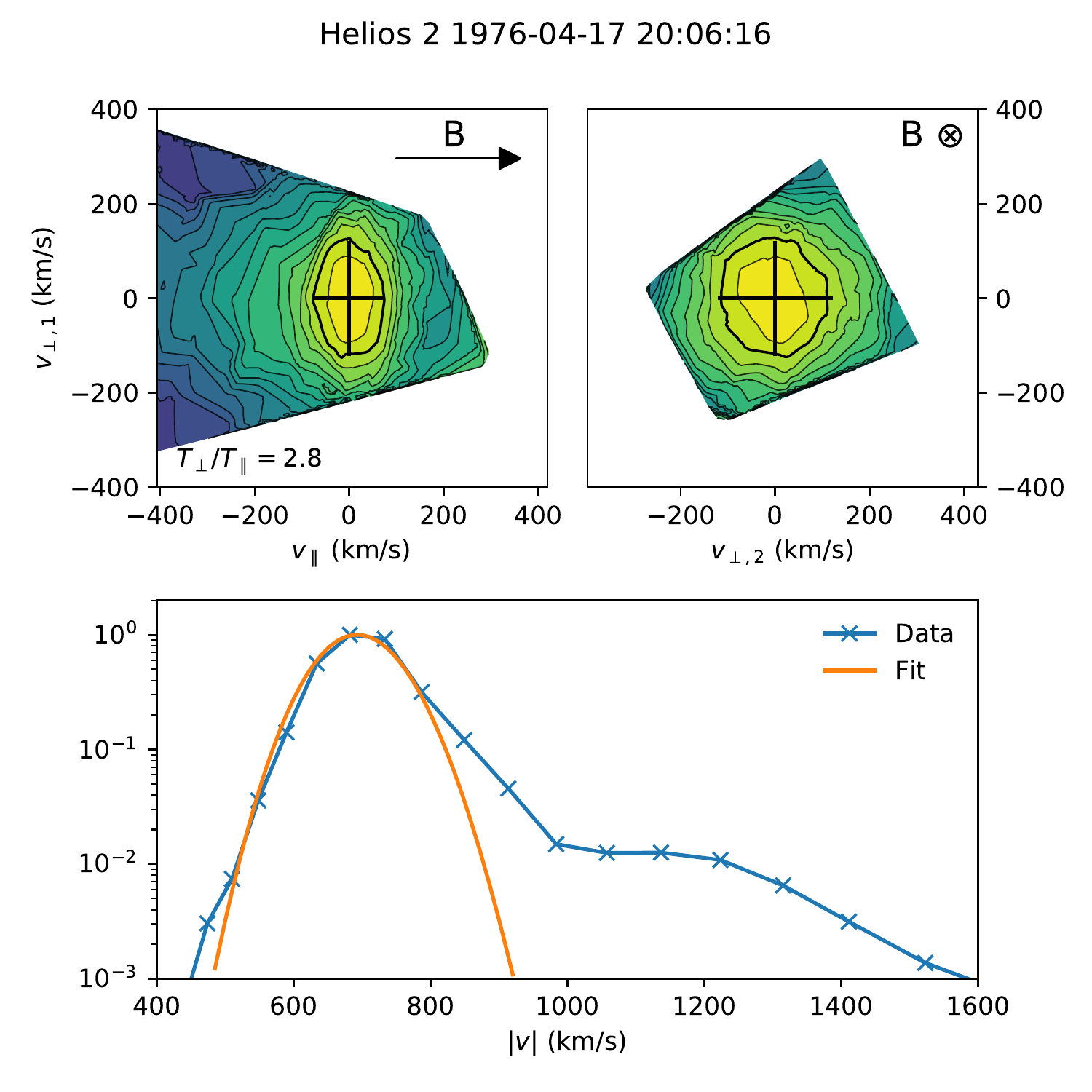}}
	\caption{Example of a fast solar wind distribution function data and corresponding fit. \newline \newline Top left panel shows a cut of the distribution function in a plane containing the local magnetic field ($\mathbf{B}$), centred at the bulk velocity. Top right panel shows a cut in the plan perpendicular to $\mathbf{B}$, also centred at the bulk velocity. In both panels, contours are spaced logarithmically and the fitted thermal speeds in each direction are shown with black crosses. The $1/e$ contour is highlighted in red, which is located one thermal width away from the centre for a bi-Maxwellian. \newline \newline Bottom panel shows the experimentally measured distribution function (blue) and fit (orange) integrated over all solid angles, and normalised to the distribution function peak.}
	\label{fig:distribution fit fast}
\end{figure}

\begin{figure}
	\centerline{\includegraphics[width=0.8\textwidth]{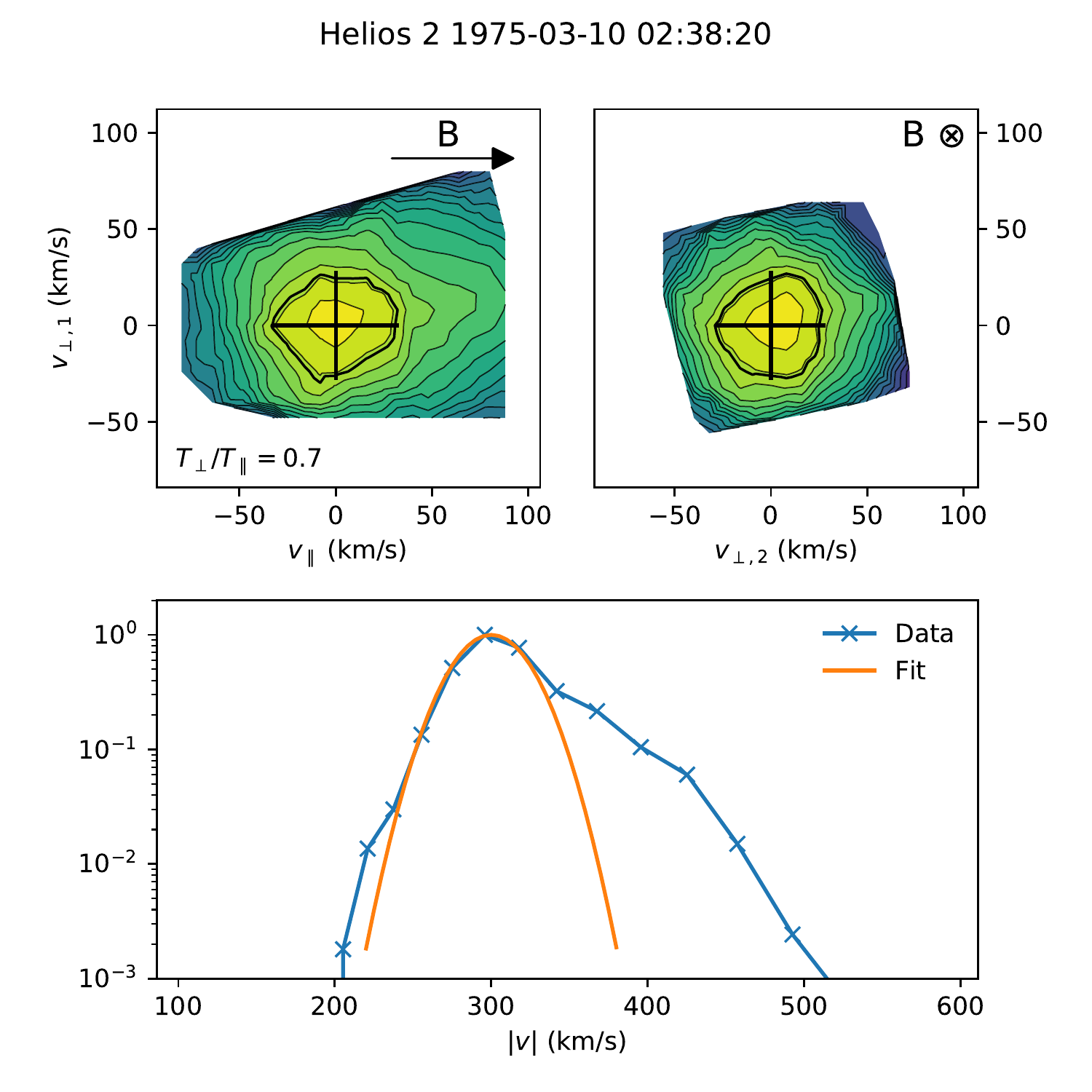}}
	\caption{Example of a slow solar wind distribution function data and corresponding fit. \newline \newline Top left panel shows a cut of the distribution function in a plane containing the local magnetic field ($\mathbf{B}$), centred at the bulk velocity. Top right panel shows a cut in the plan perpendicular to $\mathbf{B}$, also centred at the bulk velocity. In both panels, contours are spaced logarithmically and the fitted thermal speeds in each direction are shown with black crosses. The $1/e$ contour is highlighted in red, which is located one thermal width away from the centre for a bi-Maxwellian. \newline \newline Bottom panel shows the experimentally measured distribution function (blue) and fit (orange) integrated over all solid angles, and normalised to the distribution function peak.}
	\label{fig:distribution fit slow}
\end{figure}

Each experimentally measured distribution function was fitted with a bi-Maxwellian distribution function using the following process:

\begin{enumerate}
	\item If magnetic field data was available from one of the magnetic field instruments, an average magnetic field ($\mathbf{B}$) was calculated from individual measurements that fell between the time of the first and last non-zero measurements in each individual distribution function. The distribution function was then rotated into a frame aligned with $\mathbf{B}$.  This gave the rotated distribution function $f_{data} \left ( v_{\parallel}, v_{\perp 1}, v_{\perp 2} \right )$, where $v_{\parallel}$ is the direction parallel to $\mathbf{B}$ and $v_{\perp 1,2}$ are two orthogonal directions to $\mathbf{B}$ in velocity space.
	\item The following 3D bi-Maxwellian function  was fitted to the data (fit parameters underlined):
		\begin{equation}
		f_{fit} \left ( v_{\parallel}, v_{\perp 1}, v_{\perp 2} \right ) = \underline{A} \cdot \exp - \left \{ \left ( \frac{v_{\parallel} - \underline{u}_{\parallel}}{\underline{w}_{\parallel}} \right )^{2} + \left ( \frac{v_{\perp 1} - \underline{u}_{\perp 1}}{\underline{w}_{\perp}} \right )^{2} + \left ( \frac{v_{\perp 2} - \underline{u}_{\perp 2}}{\underline{w}_{\perp}} \right )^{2} \right \}
		\label{eq:bimax}
	\end{equation}
	The 6 fit parameters were amplitude ($\underline{A}$), 3 bulk velocity components ($\underline{u}_{\parallel}$, $\underline{u}_{\perp 1}$, $\underline{u}_{\perp 2}$), and 2 thermal speeds ($\underline{w}_{\perp}$, $\underline{w}_{\parallel}$). A $_\parallel$ subscript indicates a quantity parallel to $\mathbf{B}$, and a $_\perp$ subscript a quantity perpendicular to $\mathbf{B}$. The fitting was done using a least squares minimisation of the cost function
\begin{equation}
	C = \sum \left (f_{fit} - f_{data} \right )^{2}
\end{equation}
where $f_{data}$ was the experimentally measured distribution function, and the sum was taken over all velocity space points in $f_{data}$. Note that the fitting was \emph{not} done in logarithmic space\footnote{i.e.. minimising $\sum \log \left |f_{fit} - f_{data} \right |$}. This means the fitting was relatively insensitive to the tails of the distribution function, which was required to avoid the lower amplitude proton beam influencing the fit to the proton core (see figures \ref{fig:distribution fit fast} and \ref{fig:distribution fit slow} for a visual demonstration of this).
	\item The number density was calculated from
	\begin{equation}
		n = \underline{A} \cdot \pi^{3/2}\underline{w}_{\perp}\underline{w}_{\perp}\underline{w}_{\parallel}
	\end{equation}
	the two temperatures from
	\begin{equation}
		T_{\perp / \parallel} = \frac{m_{p}\underline{w}_{\perp / \parallel}^{2}}{2k_{B}}
	\end{equation}
	and the bulk velocity fitted in the field aligned frame was rotated back to the RTN instrument frame of reference to give $(v_{r}, v_{t}, v_{n})$.
\end{enumerate}

If no magnetic field values were available for any individual distribution function, it was still possible to locate the peak of the distribution, but the rotational symmetry axis of the bi-Maxwellian could not be determined. In this case the fitting still took place in the instrument (non-rotated) frame of reference, but only the velocity component values were kept and thermal speeds and number density were discarded.

If the magnetic field direction varies significantly during the time it takes to measure a distribution function, the distribution is `smeared' in the perpendicular direction, causing an overestimate of the field perpendicular temperature and number density \citep{Verscharen2011}. If any two of the magnetic field vector measurements measured during the 32 seconds the plasma instrument took to measure a full distribution were more than 90 degrees apart, the number density and temperatures were considered unreliable and not retained.

Figures \ref{fig:distribution fit fast}  and \ref{fig:distribution fit slow} show examples of 2D cuts of the original distribution functions along with bi-Maxwellian fits in both the fast and slow solar wind. Out of a total of 2,216,195 original distribution functions, 1,869,275 were successfully fit with magnetic field values (providing, density, velocity, and temperatures), and a further 227,436 were fitted without magnetic field values (providing only velocity).

\section{Comparison between \texttt{moment} and \texttt{corefit} datasets}
\label{sec:comparison}
\begin{figure}
	\centerline{\includegraphics[width=\textwidth]{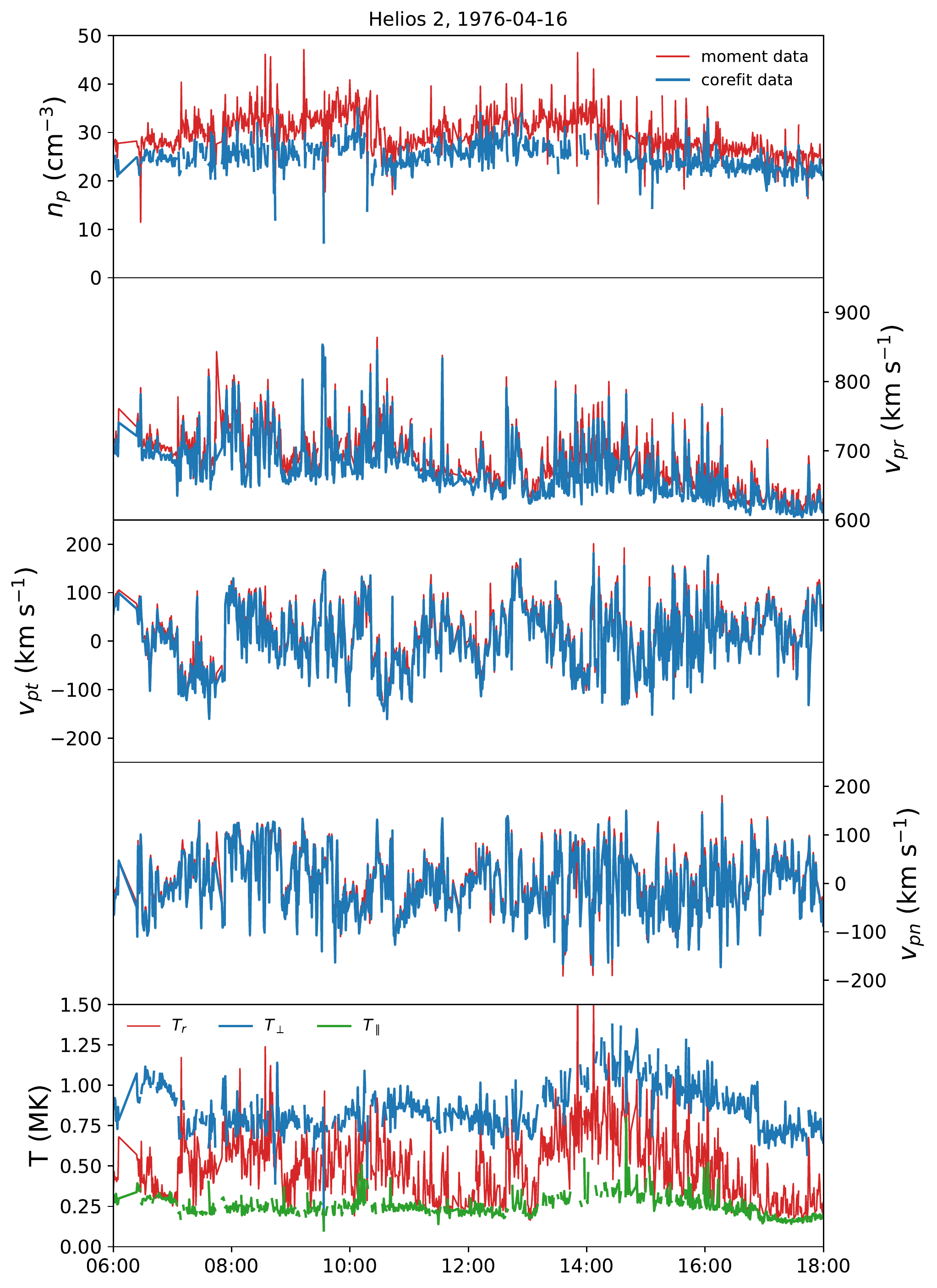}}
	\caption{A 12 hour timeseries comparing the existing and new data. \texttt{moment} data is plotted in red, and \texttt{corefit} data in blue and green. From top to bottom, proton number density, velocity components in an RTN coordinate system, and temperatures.}
	\label{fig:timeseries}
\end{figure}

Figure \ref{fig:timeseries} shows half a day of data comparing the already available \texttt{moment} dataset and the new \texttt{corefit} dataset described in section \ref{sec:processing}. The main differences between the parameters in each dataset are discussed in the following sections.

\subsection{Number density}
The number density in the \texttt{moment} data set contains contributions from the proton beam, so is systematically higher than the \texttt{corefit} number density. The difference is typically around 20\%, but can be as high as 50\% at times. A time series comparison is shown in the top panel of figure \ref{fig:timeseries}.

\subsection{Velocity}
The \texttt{moment} radial component of velocity is typically 1\% larger than the \texttt{corefit} radial velocity component, due to the presence of the proton beam.  The tangential and normal components are not affected by this and the two data sets contain very similar values. A time series comparison is shown in panels 2-4 of figure \ref{fig:timeseries}.

\subsection{Temperature}
The \texttt{moment} data set contains only one proton temperature value. This value was calculated from the reduced 1D distribution function (the 3D distribution function integrated over all solid angles), and is the projection of the numerical temperature tensor along the radial direction, which means it contains variable contributions from the true parallel and perpendicular temperatures of the protons depending on the local orientation of the magnetic field to the radial direction. The perpendicular and parallel temperatures in the \texttt{corefit} are not a function the magnetic field direction, and therefore provide a more meaningful characterisation of the true distribution function. A time series comparison is shown in the bottom panel of figure \ref{fig:timeseries}. The \texttt{corefit} total temperature, which can be calculated from the parallel and perpendicular temperatures via.
$
	T = \left ( 2T_{\perp} + T_{\parallel} \right ) / 3
$
is therefore a more accurate characterisation of the average temperature compared to the \texttt{moment} dataset. The \texttt{moment} temperature is typically 5\%, higher than the \texttt{corefit} total temperature, but the difference is highly variable and ranges from 100\% higher to 50\% lower.

\section{New radial trends}
\label{sec:trends}
\begin{figure}
	\centering
	\includegraphics[width=\columnwidth]{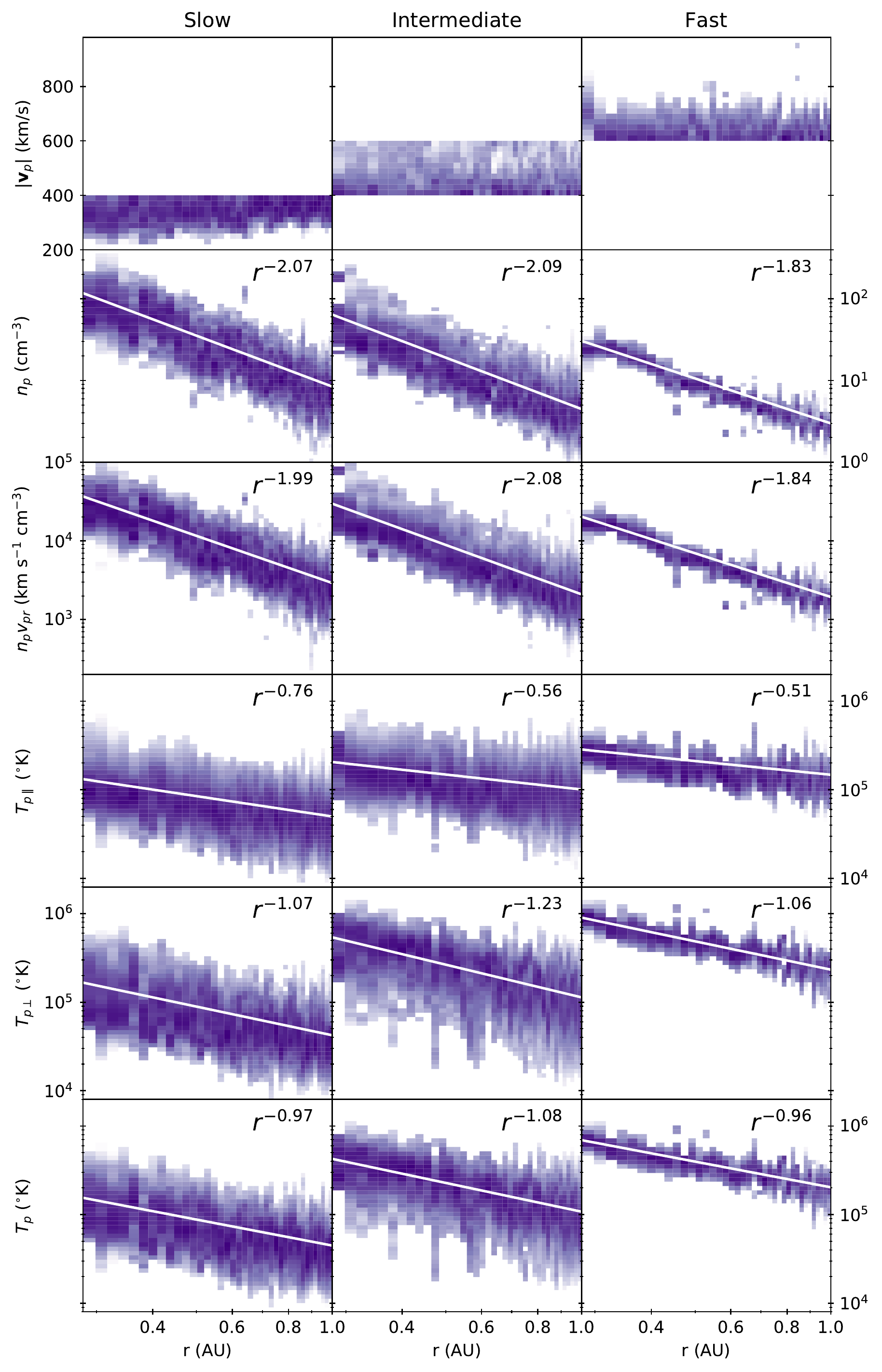}
	\caption{Radial trends of the proton core population. Histograms bins with counts greater than 100 were retained, and then normalised such that the bin values in each column sum to 1. White lines are power law fits of equation \ref{eq:fit}. Values of $A$ and $\gamma$ for each parameter are listed in table \ref{tab:fits}.}
	\label{fig:radial trends}
\end{figure}

In order to present the radial variation of parameters with distances, the data were split into slow $( | \mathbf{v}_{p} | < 400$ km/s), intermediate (400 km/s $< \left | \mathbf{v}_{p} \right | <$ 600 km/s), and fast solar wind $( | \mathbf{v}_{p} | > 600$ km/s). The radial dependence of each variable was parameterised by fitting a power law of the form
\begin{equation}
	f (r ) = A\left ( \frac{r}{r_{0}} \right )^{-\gamma}
	\label{eq:fit}
\end{equation}
to the data between 0.3 AU and 1 AU, with $r_{0}$ = 1 AU, and $A$ and $\gamma$ as two the fit parameters. 2D histograms of the variables as a function of radial distance along with the fits are shown in figure \ref{fig:radial trends}. The fitted values of $A$ and $\gamma$ for each variable and category of solar wind are reported in table \ref{tab:fits}.

\begin{table}
	\begin{tabular*}{\textwidth}{ c  c  c  c  c  c  c }
		\hline
				&  \multicolumn{3}{c}{$\gamma$}	& \multicolumn{3}{c}{$A$}		\\ 
					& Slow	& Intermediate  & Fast	& Slow				& Intermediate			& Fast				\\ \hline
		$n_{p}$		& 2.07	& 2.09		& 1.83	& 8.44 cm$^{-3}$		& 4.47 cm$^{-3}$		& 2.98 cm$^{-3}$		\\
		$n_{p} v_{pr}$	& 1.99	& 2.08		& 1.84 	& 2910 cm$^{-2}$s$^{-1}$	& 2092 cm$^{-2}$s$^{-1}$	& 1936 cm$^{-2}$s$^{-1}$	\\ 
		$T_{p\parallel}$& 0.76	& 0.56		& 0.51	& 0.0500 MK			& 0.101 MK			& 0.148 MK			\\ 
		$T_{p\perp}$	& 1.07	& 1.23		& 1.06	& 0.0423 MK			& 0.113 MK			& 0.233 MK			\\ 
		$T_{p}$		& 0.97	& 1.08		& 0.96	& 0.0447 MK			& 0.108 MK			& 0.203 MK			\\ \hline
	\end{tabular*}
	\caption{Results of power law fits as a function of radial distance. Fits are parameterised by equation \ref{eq:fit}. $A$ is the 1 AU intercept, and $\gamma$ is the power law exponent. See figure \ref{fig:radial trends} for a visual comparison of the fitted curves and underlying data.}
	\label{tab:fits}
\end{table}

From the new radial trends the following well known results are reproduced:
\begin{itemize}
	\item The number density in the slow solar wind is larger and more variable than in the fast solar wind.
	\item The number density decreases faster than a simple 1/$r^{2}$ constant speed decrease in the slow solar wind ($\gamma$ = 2.07) and slower than simple radial expansion in the fast solar wind ($\gamma =1.83$). This is most likely due to slow solar wind accelerating and fast solar wind being decelerating between 0.3 AU and 1 AU.
	\item The radial flux almost follows a $1/r^{2}$ decrease in the slow solar wind ($\gamma = 1.99$), but decreases slower in the fast solar wind ($\gamma = 1.84$).
	\item The slowest solar wind at 0.3 AU ($\sim$ 200 km/s) is accelerated up to a larger minimum ($\sim$ 250 km/s) by 1 AU.
\end{itemize}
In addition, the trends successfully reproduce a number of features in the radial evolution of temperatures initially observed by \cite{Marsch1982b}:
\begin{itemize}
	\item Both the $T_{p\perp}$ and $T_{p\parallel}$ are higher and less variable in the fast solar wind.
	\item $T_{p\perp}$ decreases faster with radial distance than $T_{p\parallel}$.
	\item $T_{p\perp}$ decreases faster with radial distance in fast solar wind (compared to slow wind), whereas $T_{p\parallel}$ decreases faster with radial distance in slow solar wind (compared to fast wind). 
	\item $T_{p\perp}$ and $T_{p\parallel}$  both decrease slower than a single adiabatic prediction ($\gamma$ = 5/3).
	\item $T_{p\perp}$ decreases slower than the  \cite{Chew1956} double adiabatic prediction ($\gamma =  2$), but $T_{p\parallel}$ decreases faster than the prediction ($\gamma$ = 0).
\end{itemize}

\cite{Marsch1982b} and \cite{Hellinger2011, Hellinger2013a} have previously performed similar analyses of the Helios data to extract the parallel and perpendicular temperatures, using numerical moments of the distribution function instead of analytical fits. This means that they did not separate out the contributions from the proton core and beam. The presence of a beam lead to larger $T_{p\parallel}$ values in both sets of data compared to ours.

Finally, we note that combining data from a wide range of times and locations into single radial fits means that the data are not sampling how a single parcel of plasma evolves as it propagates radially outwards. Nonetheless this type of analysis is useful for indicating average behaviour of the solar wind.

\section{Conclusion}
We have presented the method and results of a complete reprocessing of the original Helios solar wind ion distribution functions, measured between 0.3 AU and 1 AU. The resulting dataset has been made freely available on the Helios data archive (\url{http://helios-data.ssl.berkeley.edu/}) for other researchers to use, and the code used to fit the distributions functions has also been made available, making the dataset reproducible. The new data provides a benchmark of how the proton core evolves in the inner heliosphere. This dataset forms an important resource to which in-situ data from the upcoming Parker Solar Probe and Solar Orbiter missions will be compared against to study variations of the solar wind on decadal timescales.


\begin{acknowledgements}
This work has built upon efforts by all members of the Helios data archive team (\url{http://helios-data.ssl.berkeley.edu/team-members/}) to make the Helios data publicly available to the space physics community.
\newline \newline
D.~Stansby is supported by STFC studentship ST/N504336/1, and thanks Denise Perrone for helpful discusisons. Work at UC Berkeley was supported by NASA grants NNX14AQ89G and 80NSSC18K0370. T.~S.~Horbury is supported by STFC grant ST/N000692/1. 
\newline \newline
A copy of this new Helios proton core data set, along with the source code used to generate it, is available online and citeable at \url{https://zenodo.org/record/1009506} \citep{Stansby2017c}. 
\newline \newline
Figures were produced using Matplotlib v2.2.2 \citep{Hunter2007, Droettboom2018}, data retrieved using HelioPy v0.5.1 \citep{Stansby2018b} and processed using astropy v3.0 \citep{TheAstropyCollaboration2018}.
\end{acknowledgements}

\bibliographystyle{spr-mp-sola}
\bibliography{/Users/dstansby/Dropbox/Physics/library}  

\begin{thebibliography}{24}
\ifx\bisbn     \undefined \def\bisbn  #1{ISBN #1}\fi
\ifx\binits    \undefined \def\binits#1{#1}\fi
\ifx\bauthor   \undefined \def\bauthor#1{#1}\fi
\ifx\batitle   \undefined \def\batitle#1{#1}\fi
\ifx\bjtitle   \undefined \def\bjtitle#1{\textit{#1}}\fi
\ifx\bvolume   \undefined \def\bvolume#1{\textbf{#1}}\fi
\ifx\byear     \undefined \def\byear#1{#1}\fi
\ifx\bissue    \undefined \def\bissue#1{#1}\fi
\ifx\bfpage    \undefined \def\bfpage#1{#1}\fi
\ifx\blpage    \undefined \def\blpage #1{#1}\fi
\ifx\burl      \undefined \def\burl#1{\textsf{#1}}\fi
\ifx\href      \undefined \def\href#1#2{\textsf{#2}}\fi
\ifx\betal     \undefined \def\betal{\textit{et al.}}\fi
\ifx\bctitle   \undefined \def\bctitle#1{#1}\fi
\ifx\beditor   \undefined \def\beditor#1{#1}\fi
\ifx\bbtitle   \undefined \def\bbtitle#1{\textit{#1}}\fi
\ifx\bedition  \undefined \def\bedition#1{#1}\fi
\ifx\bseriesno \undefined \def\bseriesno#1{\textbf{#1}}\fi
\ifx\blocation \undefined \def\blocation#1{#1}\fi
\ifx\bsertitle \undefined \def\bsertitle#1{\textit{#1}}\fi
\ifx\bsnm      \undefined \def\bsnm#1{#1}\fi
\ifx\bsuffix   \undefined \def\bsuffix#1{#1}\fi
\ifx\bparticle \undefined \def\bparticle#1{#1}\fi
\ifx\barticle  \undefined \def\barticle#1{}\fi
\ifx\binstitute  \undefined \def\binstitute#1{#1}\fi
\ifx\bpublisher  \undefined \def\bpublisher#1{#1}\fi
\ifx\doiurl    \undefined
  \def\doiurl#1{\href{http://dx.doi.org/#1}{\textsf{DOI}}}\fi
\ifx\arxivurl  \undefined
  \def\arxivurl#1{\href{http://arxiv.org/abs/#1}{\textsf{arXiv}}}\fi
\ifx\adsurl    \undefined
  \def\adsurl#1{\href{http://adsabs.harvard.edu/abs/#1}{\textsf{ADS}}}\fi
\ifx\botherref \undefined \def\botherref#1{}\fi
\ifx\url       \undefined \def\url#1{\textsf{#1}}\fi
\ifx\bchapter  \undefined \def\bchapter#1{}\fi
\ifx\bbook     \undefined \def\bbook#1{}\fi
\ifx\bcomment  \undefined \def\bcomment#1{#1}\fi
\ifx\oauthor   \undefined \def\oauthor#1{#1}\fi
\ifx\citeauthoryear \undefined\def \citeauthoryear#1{#1}\fi
\ifx\endbibitem\undefined \def\endbibitem{}\fi
\ifx\bconflocation  \undefined \def\bconflocation#1{#1} \fi

\bibitem[\protect\citeauthoryear{Bochsler}{2007}]{Bochsler2007}
\begin{barticle}
\bauthor{\bsnm{Bochsler}, \binits{P.}}:
\byear{2007},
\batitle{{Minor ions in the solar wind}}.
\bjtitle{The Astronomy and Astrophysics Review}
\bvolume{14}(\bissue{1}),
\bfpage{1}.
\doiurl{10.1007/s00159-006-0002-x}.
\burl{http://link.springer.com/10.1007/s00159-006-0002-x}.
\end{barticle}
\endbibitem

\bibitem[\protect\citeauthoryear{Chew, Goldberger, and Low}{1956}]{Chew1956}
\begin{barticle}
\bauthor{\bsnm{Chew}, \binits{G.F.}},
\bauthor{\bsnm{Goldberger}, \binits{M.L.}},
\bauthor{\bsnm{Low}, \binits{F.E.}}:
\byear{1956},
\batitle{{The Boltzmann Equation and the One-Fluid Hydromagnetic Equations in
  the Absence of Particle Collisions}}.
\bjtitle{Proceedings of the Royal Society A: Mathematical, Physical and
  Engineering Sciences}
\bvolume{236}(\bissue{1204}),
\bfpage{112}.
\doiurl{10.1098/rspa.1956.0116}.
\end{barticle}
\endbibitem

\bibitem[\protect\citeauthoryear{Droettboom
  \textit{et~al.}}{2018}]{Droettboom2018}
\begin{botherref}
\oauthor{\bsnm{Droettboom}, \binits{M.}},
\oauthor{\bsnm{Caswell}, \binits{T.A.}},
\oauthor{\bsnm{Hunter}, \binits{J.}},
\oauthor{\bsnm{Firing}, \binits{E.}},
\oauthor{\bsnm{Nielsen}, \binits{J.H.}},
\oauthor{\bsnm{Lee}, \binits{A.}},
\oauthor{\bparticle{de} \bsnm{Andrade}, \binits{E.S.}},
\oauthor{\bsnm{Varoquaux}, \binits{N.}},
\oauthor{\bsnm{Stansby}, \binits{D.}},
\oauthor{\bsnm{Root}, \binits{B.}},
\oauthor{\bsnm{Elson}, \binits{P.}},
\oauthor{\bsnm{Dale}, \binits{D.}},
\oauthor{\bsnm{Lee}, \binits{J.-J.}},
\oauthor{\bsnm{May}, \binits{R.}},
\oauthor{\bsnm{Sepp{\"{a}}nen}, \binits{J.K.}},
\oauthor{\bsnm{Klymak}, \binits{J.}},
\oauthor{\bsnm{McDougall}, \binits{D.}},
\oauthor{\bsnm{Straw}, \binits{A.}},
\oauthor{\bsnm{Hobson}, \binits{P.}},
\oauthor{\bsnm{Cgohlke}},
\oauthor{\bsnm{Yu}, \binits{T.S.}},
\oauthor{\bsnm{Ma}, \binits{E.}},
\oauthor{\bsnm{Vincent}, \binits{A.F.}},
\oauthor{\bsnm{Silvester}, \binits{S.}},
\oauthor{\bsnm{Moad}, \binits{C.}},
\oauthor{\bsnm{Katins}, \binits{J.}},
\oauthor{\bsnm{Kniazev}, \binits{N.}},
\oauthor{\bsnm{Hoffmann}, \binits{T.}},
\oauthor{\bsnm{Ariza}, \binits{F.}},
\oauthor{\bsnm{W{\"{u}}rtz}, \binits{P.}}:
2018,
matplotlib/matplotlib v2.2.2.
\doiurl{10.5281/ZENODO.1202077}.
\url{https://zenodo.org/record/1202077}.
\end{botherref}
\endbibitem

\bibitem[\protect\citeauthoryear{Feldman \textit{et~al.}}{1973}]{Feldman1973}
\begin{barticle}
\bauthor{\bsnm{Feldman}, \binits{W.C.}},
\bauthor{\bsnm{Asbridge}, \binits{J.R.}},
\bauthor{\bsnm{Bame}, \binits{S.J.}},
\bauthor{\bsnm{Montgomery}, \binits{M.D.}}:
\byear{1973},
\batitle{{Double ion streams in the solar wind}}.
\bjtitle{Journal of Geophysical Research}
\bvolume{78}(\bissue{13}),
\bfpage{2017}.
\doiurl{10.1029/JA078i013p02017}.
\burl{http://doi.wiley.com/10.1029/JA078i013p02017}.
\end{barticle}
\endbibitem

\bibitem[\protect\citeauthoryear{Fox \textit{et~al.}}{2016}]{Fox2015}
\begin{barticle}
\bauthor{\bsnm{Fox}, \binits{N.J.}},
\bauthor{\bsnm{Velli}, \binits{M.C.}},
\bauthor{\bsnm{Bale}, \binits{S.D.}},
\bauthor{\bsnm{Decker}, \binits{R.}},
\bauthor{\bsnm{Driesman}, \binits{A.}},
\bauthor{\bsnm{Howard}, \binits{R.A.}},
\bauthor{\bsnm{Kasper}, \binits{J.C.}},
\bauthor{\bsnm{Kinnison}, \binits{J.}},
\bauthor{\bsnm{Kusterer}, \binits{M.}},
\bauthor{\bsnm{Lario}, \binits{D.}},
\bauthor{\bsnm{Lockwood}, \binits{M.K.}},
\bauthor{\bsnm{McComas}, \binits{D.J.}},
\bauthor{\bsnm{Raouafi}, \binits{N.E.}},
\bauthor{\bsnm{Szabo}, \binits{A.}}:
\byear{2016},
\batitle{{The Solar Probe Plus Mission: Humanity?s First Visit to Our Star}}.
\bjtitle{Space Science Reviews}
\bvolume{204}(\bissue{1-4}),
\bfpage{7}.
\doiurl{10.1007/s11214-015-0211-6}.
\burl{http://link.springer.com/10.1007/s11214-015-0211-6}.
\end{barticle}
\endbibitem

\bibitem[\protect\citeauthoryear{Hellinger
  \textit{et~al.}}{2011}]{Hellinger2011}
\begin{barticle}
\bauthor{\bsnm{Hellinger}, \binits{P.}},
\bauthor{\bsnm{Matteini}, \binits{L.}},
\bauthor{\bsnm{{\v{S}}tver{\'{a}}k}, \binits{{\v{S}}.}},
\bauthor{\bsnm{Tr{\'{a}}vn{\'{i}}{\v{c}}ek}, \binits{P.M.}},
\bauthor{\bsnm{Marsch}, \binits{E.}}:
\byear{2011},
\batitle{{Heating and cooling of protons in the fast solar wind between 0.3 and
  1 AU: Helios revisited}}.
\bjtitle{Journal of Geophysical Research: Space Physics}
\bvolume{116}(\bissue{A9}),
\bfpage{n/a}.
\doiurl{10.1029/2011JA016674}.
\burl{http://doi.wiley.com/10.1029/2011JA016674}.
\end{barticle}
\endbibitem

\bibitem[\protect\citeauthoryear{Hellinger
  \textit{et~al.}}{2013}]{Hellinger2013a}
\begin{barticle}
\bauthor{\bsnm{Hellinger}, \binits{P.}},
\bauthor{\bsnm{Tr{\'{a}}vn{\'{i}}{\v{c}}ek}, \binits{P.M.}},
\bauthor{\bsnm{{\v{S}}tver{\'{a}}k}, \binits{{\v{S}}.}},
\bauthor{\bsnm{Matteini}, \binits{L.}},
\bauthor{\bsnm{Velli}, \binits{M.}}:
\byear{2013},
\batitle{{Proton thermal energetics in the solar wind: Helios reloaded}}.
\bjtitle{Journal of Geophysical Research: Space Physics}
\bvolume{118}(\bissue{4}),
\bfpage{1351}.
\doiurl{10.1002/jgra.50107}.
\burl{http://doi.wiley.com/10.1002/jgra.50107}.
\end{barticle}
\endbibitem

\bibitem[\protect\citeauthoryear{Hunter}{2007}]{Hunter2007}
\begin{barticle}
\bauthor{\bsnm{Hunter}, \binits{J.D.}}:
\byear{2007},
\batitle{{Matplotlib: A 2D Graphics Environment}}.
\bjtitle{Computing in Science {\&} Engineering}
\bvolume{9}(\bissue{3}),
\bfpage{90}.
\doiurl{10.1109/MCSE.2007.55}.
\burl{http://ieeexplore.ieee.org/document/4160265/}.
\end{barticle}
\endbibitem

\bibitem[\protect\citeauthoryear{Marsch, Ao, and Tu}{2004}]{Marsch2004}
\begin{barticle}
\bauthor{\bsnm{Marsch}, \binits{E.}},
\bauthor{\bsnm{Ao}, \binits{X.-Z.}},
\bauthor{\bsnm{Tu}, \binits{C.-Y.}}:
\byear{2004},
\batitle{{On the temperature anisotropy of the core part of the proton velocity
  distribution function in the solar wind}}.
\bjtitle{Journal of Geophysical Research}
\bvolume{109}(\bissue{A4}),
\bfpage{A04102}.
\doiurl{10.1029/2003JA010330}.
\burl{http://doi.wiley.com/10.1029/2003JA010330}.
\end{barticle}
\endbibitem

\bibitem[\protect\citeauthoryear{Marsch \textit{et~al.}}{1982a}]{Marsch1982c}
\begin{barticle}
\bauthor{\bsnm{Marsch}, \binits{E.}},
\bauthor{\bsnm{M{\"{u}}hlh{\"{a}}user}, \binits{K.-H.}},
\bauthor{\bsnm{Rosenbauer}, \binits{H.}},
\bauthor{\bsnm{Schwenn}, \binits{R.}},
\bauthor{\bsnm{Neubauer}, \binits{F.M.}}:
\byear{1982}a,
\batitle{{Solar wind helium ions: Observations of the Helios solar probes
  between 0.3 and 1 AU}}.
\bjtitle{Journal of Geophysical Research}
\bvolume{87}(\bissue{A1}),
\bfpage{35}.
\doiurl{10.1029/JA087iA01p00035}.
\burl{http://doi.wiley.com/10.1029/JA087iA01p00035}.
\end{barticle}
\endbibitem

\bibitem[\protect\citeauthoryear{Marsch \textit{et~al.}}{1982b}]{Marsch1982b}
\begin{barticle}
\bauthor{\bsnm{Marsch}, \binits{E.}},
\bauthor{\bsnm{M{\"{u}}hlh{\"{a}}user}, \binits{K.-H.}},
\bauthor{\bsnm{Schwenn}, \binits{R.}},
\bauthor{\bsnm{Rosenbauer}, \binits{H.}},
\bauthor{\bsnm{Pilipp}, \binits{W.}},
\bauthor{\bsnm{Neubauer}, \binits{F.M.}}:
\byear{1982}b,
\batitle{{Solar wind protons: Three-dimensional velocity distributions and
  derived plasma parameters measured between 0.3 and 1 AU}}.
\bjtitle{Journal of Geophysical Research}
\bvolume{87}(\bissue{A1}),
\bfpage{52}.
\doiurl{10.1029/JA087iA01p00052}.
\burl{http://doi.wiley.com/10.1029/JA087iA01p00052}.
\end{barticle}
\endbibitem

\bibitem[\protect\citeauthoryear{Matteini \textit{et~al.}}{2007}]{Matteini2007}
\begin{barticle}
\bauthor{\bsnm{Matteini}, \binits{L.}},
\bauthor{\bsnm{Landi}, \binits{S.}},
\bauthor{\bsnm{Hellinger}, \binits{P.}},
\bauthor{\bsnm{Pantellini}, \binits{F.}},
\bauthor{\bsnm{Maksimovic}, \binits{M.}},
\bauthor{\bsnm{Velli}, \binits{M.}},
\bauthor{\bsnm{Goldstein}, \binits{B.E.}},
\bauthor{\bsnm{Marsch}, \binits{E.}}:
\byear{2007},
\batitle{{Evolution of the solar wind proton temperature anisotropy from 0.3 to
  2.5 AU}}.
\bjtitle{Geophysical Research Letters}
\bvolume{34}(\bissue{20}),
\bfpage{L20105}.
\doiurl{10.1029/2007GL030920}.
\burl{http://doi.wiley.com/10.1029/2007GL030920}.
\end{barticle}
\endbibitem

\bibitem[\protect\citeauthoryear{M{\"{u}}ller
  \textit{et~al.}}{2013}]{Muller2013}
\begin{barticle}
\bauthor{\bsnm{M{\"{u}}ller}, \binits{D.}},
\bauthor{\bsnm{Marsden}, \binits{R.G.}},
\bauthor{\bsnm{{St. Cyr}}, \binits{O.C.}},
\bauthor{\bsnm{Gilbert}, \binits{H.R.}}:
\byear{2013},
\batitle{{Solar Orbiter}}.
\bjtitle{Solar Physics}
\bvolume{285}(\bissue{1-2}),
\bfpage{25}.
\doiurl{10.1007/s11207-012-0085-7}.
\burl{http://link.springer.com/10.1007/s11207-012-0085-7}.
\end{barticle}
\endbibitem

\bibitem[\protect\citeauthoryear{Musmann \textit{et~al.}}{1975}]{Musmann1975}
\begin{barticle}
\bauthor{\bsnm{Musmann}, \binits{G.}},
\bauthor{\bsnm{Neubauer}, \binits{F.M.}},
\bauthor{\bsnm{Maier}, \binits{A.}},
\bauthor{\bsnm{Lammers}, \binits{E.}}:
\byear{1975},
\batitle{{The Foerstersonden magnetic field experiment E2}}.
\bjtitle{Raumfahrtforschung}
\bvolume{19},
\bfpage{232}.
\burl{https://ui.adsabs.harvard.edu/{\#}abs/1975RF.....19..232M/abstract}.
\end{barticle}
\endbibitem

\bibitem[\protect\citeauthoryear{Neugebauer and Snyder}{1962}]{Neugebauer1962}
\begin{barticle}
\bauthor{\bsnm{Neugebauer}, \binits{M.}},
\bauthor{\bsnm{Snyder}, \binits{C.W.}}:
\byear{1962},
\batitle{{Solar Plasma Experiment.}}
\bjtitle{Science (New York, N.Y.)}
\bvolume{138}(\bissue{3545}),
\bfpage{1095}.
\doiurl{10.1126/science.138.3545.1095-a}.
\burl{http://www.ncbi.nlm.nih.gov/pubmed/17772963}.
\end{barticle}
\endbibitem

\bibitem[\protect\citeauthoryear{Pilipp \textit{et~al.}}{1987}]{Pilipp1987}
\begin{barticle}
\bauthor{\bsnm{Pilipp}, \binits{W.G.}},
\bauthor{\bsnm{Miggenrieder}, \binits{H.}},
\bauthor{\bsnm{Montgomery}, \binits{M.D.}},
\bauthor{\bsnm{M{\"{u}}hlh{\"{a}}user}, \binits{K.-H.}},
\bauthor{\bsnm{Rosenbauer}, \binits{H.}},
\bauthor{\bsnm{Schwenn}, \binits{R.}}:
\byear{1987},
\batitle{{Characteristics of electron velocity distribution functions in the
  solar wind derived from the Helios Plasma Experiment}}.
\bjtitle{Journal of Geophysical Research}
\bvolume{92}(\bissue{A2}),
\bfpage{1075}.
\doiurl{10.1029/JA092iA02p01075}.
\burl{http://doi.wiley.com/10.1029/JA092iA02p01075}.
\end{barticle}
\endbibitem

\bibitem[\protect\citeauthoryear{Porsche}{1977}]{PORSCHE1977}
\begin{barticle}
\bauthor{\bsnm{Porsche}, \binits{H.}}:
\byear{1977},
\batitle{{GENERAL ASPECTS OF MISSION HELIOS-1 AND MISSION HELIOS-2 -
  INTRODUCTION TO A SPECIAL ISSUE ON INITIAL SCIENTIFIC RESULTS OF HELIOS
  MISSION}}.
\bjtitle{Journal of Geophysics}
\bvolume{42}(\bissue{6}),
\bfpage{551}.
\end{barticle}
\endbibitem

\bibitem[\protect\citeauthoryear{Rosenbauer
  \textit{et~al.}}{1981}]{Rosenbauer2018}
\begin{botherref}
\oauthor{\bsnm{Rosenbauer}, \binits{H.}},
\oauthor{\bsnm{Schwenn}, \binits{R.}},
\oauthor{\bsnm{Miggenrieder}, \binits{H.}},
\oauthor{\bsnm{Meyer}, \binits{B.}},
\oauthor{\bsnm{Gr{\"{u}}ndwaldt}, \binits{H.}},
\oauthor{\bsnm{M{\"{u}}hlh{\"{a}}user}, \binits{K.-H.}},
\oauthor{\bsnm{Pellkofer}, \binits{H.}},
\oauthor{\bsnm{Wolfe}, \binits{J.H.}}:
1981,
{Helios E1 (plasma) instrument technical document}.
Technical report.
\doiurl{10.5281/ZENODO.1240455}.
\url{https://zenodo.org/record/1240455}.
\end{botherref}
\endbibitem

\bibitem[\protect\citeauthoryear{Scearce \textit{et~al.}}{1975}]{Scearce1975}
\begin{barticle}
\bauthor{\bsnm{Scearce}, \binits{C.}},
\bauthor{\bsnm{Cantarano}, \binits{S.}},
\bauthor{\bsnm{Ness}, \binits{N.}},
\bauthor{\bsnm{Mariani}, \binits{F.}},
\bauthor{\bsnm{Terenzi}, \binits{R.}},
\bauthor{\bsnm{Burlaga}, \binits{L.}}:
\byear{1975},
\batitle{{The Rome-GSFC magnetic field experiment for Helios A and B (E3).}}
\bjtitle{Raumfahrtforschung}
\bvolume{19},
\bfpage{237}.
\burl{https://ui.adsabs.harvard.edu/{\#}abs/1975RF.....19..237S/abstract}.
\end{barticle}
\endbibitem

\bibitem[\protect\citeauthoryear{Schwenn, Rosenbauer, and
  Miggenrieder}{1975}]{Schwenn1975}
\begin{barticle}
\bauthor{\bsnm{Schwenn}, \binits{R.}},
\bauthor{\bsnm{Rosenbauer}, \binits{H.}},
\bauthor{\bsnm{Miggenrieder}, \binits{H.}}:
\byear{1975},
\batitle{{The plasma experiment on board Helios E1}}.
\bjtitle{Raumfahrtforschung}
\bvolume{19},
\bfpage{226}.
\burl{http://adsabs.harvard.edu/abs/1975RF.....19..226S}.
\end{barticle}
\endbibitem

\bibitem[\protect\citeauthoryear{Stansby}{2017}]{Stansby2017c}
\begin{botherref}
\oauthor{\bsnm{Stansby}, \binits{D.}}:
2017,
{Helios proton core parameter dataset}.
\textit{http://doi.org/10.5281/zenodo.1009506}.
\doiurl{10.5281/zenodo.1009506}.
\url{https://zenodo.org/record/1009506}.
\end{botherref}
\endbibitem

\bibitem[\protect\citeauthoryear{Stansby, Yatharth, and
  Shaw}{2018}]{Stansby2018b}
\begin{botherref}
\oauthor{\bsnm{Stansby}, \binits{D.}},
\oauthor{\bsnm{Yatharth}},
\oauthor{\bsnm{Shaw}, \binits{S.}}:
2018,
{heliopython/heliopy: HelioPy 0.5.1}.
\doiurl{10.5281/ZENODO.1237716}.
\url{https://zenodo.org/record/1237716}.
\end{botherref}
\endbibitem

\bibitem[\protect\citeauthoryear{{The Astropy Collaboration}
  \textit{et~al.}}{2018}]{TheAstropyCollaboration2018}
\begin{botherref}
\oauthor{\bsnm{{The Astropy Collaboration}}, \binits{T.A.}},
\oauthor{\bsnm{Price-Whelan}, \binits{A.M.}},
\oauthor{\bsnm{Sipőcz}, \binits{B.M.}},
\oauthor{\bsnm{G{\"{u}}nther}, \binits{H.M.}},
\oauthor{\bsnm{Lim}, \binits{P.L.}},
\oauthor{\bsnm{Crawford}, \binits{S.M.}},
\oauthor{\bsnm{Conseil}, \binits{S.}},
\oauthor{\bsnm{Shupe}, \binits{D.L.}},
\oauthor{\bsnm{Craig}, \binits{M.W.}},
\oauthor{\bsnm{Dencheva}, \binits{N.}},
\oauthor{\bsnm{Ginsburg}, \binits{A.}},
\oauthor{\bsnm{VanderPlas}, \binits{J.T.}},
\oauthor{\bsnm{Bradley}, \binits{L.D.}},
\oauthor{\bsnm{P{\'{e}}rez-Su{\'{a}}rez}, \binits{D.}},
\oauthor{\bparticle{de} \bsnm{Val-Borro}, \binits{M.}},
\oauthor{\bsnm{Aldcroft}, \binits{T.L.}},
\oauthor{\bsnm{Cruz}, \binits{K.L.}},
\oauthor{\bsnm{Robitaille}, \binits{T.P.}},
\oauthor{\bsnm{Tollerud}, \binits{E.J.}},
\oauthor{\bsnm{Ardelean}, \binits{C.}},
\oauthor{\bsnm{Babej}, \binits{T.}},
\oauthor{\bsnm{Bachetti}, \binits{M.}},
\oauthor{\bsnm{Bakanov}, \binits{A.V.}},
\oauthor{\bsnm{Bamford}, \binits{S.P.}},
\oauthor{\bsnm{Barentsen}, \binits{G.}},
\oauthor{\bsnm{Barmby}, \binits{P.}},
\oauthor{\bsnm{Baumbach}, \binits{A.}},
\oauthor{\bsnm{Berry}, \binits{K.L.}},
\oauthor{\bsnm{Biscani}, \binits{F.}},
\oauthor{\bsnm{Boquien}, \binits{M.}},
\oauthor{\bsnm{Bostroem}, \binits{K.A.}},
\oauthor{\bsnm{Bouma}, \binits{L.G.}},
\oauthor{\bsnm{Brammer}, \binits{G.B.}},
\oauthor{\bsnm{Bray}, \binits{E.M.}},
\oauthor{\bsnm{Breytenbach}, \binits{H.}},
\oauthor{\bsnm{Buddelmeijer}, \binits{H.}},
\oauthor{\bsnm{Burke}, \binits{D.J.}},
\oauthor{\bsnm{Calderone}, \binits{G.}},
\oauthor{\bsnm{Rodr{\'{i}}guez}, \binits{J.L.C.}},
\oauthor{\bsnm{Cara}, \binits{M.}},
\oauthor{\bsnm{Cardoso}, \binits{J.V.M.}},
\oauthor{\bsnm{Cheedella}, \binits{S.}},
\oauthor{\bsnm{Copin}, \binits{Y.}},
\oauthor{\bsnm{Crichton}, \binits{D.}},
\oauthor{\bsnm{D{\'{A}}vella}, \binits{D.}},
\oauthor{\bsnm{Deil}, \binits{C.}},
\oauthor{\bsnm{Depagne}, \binits{{\'{E}}.}},
\oauthor{\bsnm{Dietrich}, \binits{J.P.}},
\oauthor{\bsnm{Donath}, \binits{A.}},
\oauthor{\bsnm{Droettboom}, \binits{M.}},
\oauthor{\bsnm{Earl}, \binits{N.}},
\oauthor{\bsnm{Erben}, \binits{T.}},
\oauthor{\bsnm{Fabbro}, \binits{S.}},
\oauthor{\bsnm{Ferreira}, \binits{L.A.}},
\oauthor{\bsnm{Finethy}, \binits{T.}},
\oauthor{\bsnm{Fox}, \binits{R.T.}},
\oauthor{\bsnm{Garrison}, \binits{L.H.}},
\oauthor{\bsnm{Gibbons}, \binits{S.L.J.}},
\oauthor{\bsnm{Goldstein}, \binits{D.A.}},
\oauthor{\bsnm{Gommers}, \binits{R.}},
\oauthor{\bsnm{Greco}, \binits{J.P.}},
\oauthor{\bsnm{Greenfield}, \binits{P.}},
\oauthor{\bsnm{Groener}, \binits{A.M.}},
\oauthor{\bsnm{Grollier}, \binits{F.}},
\oauthor{\bsnm{Hagen}, \binits{A.}},
\oauthor{\bsnm{Hirst}, \binits{P.}},
\oauthor{\bsnm{Homeier}, \binits{D.}},
\oauthor{\bsnm{Horton}, \binits{A.J.}},
\oauthor{\bsnm{Hosseinzadeh}, \binits{G.}},
\oauthor{\bsnm{Hu}, \binits{L.}},
\oauthor{\bsnm{Hunkeler}, \binits{J.S.}},
\oauthor{\bsnm{Ivezi{\'{c}}}, \binits{{\v{Z}}.}},
\oauthor{\bsnm{Jain}, \binits{A.}},
\oauthor{\bsnm{Jenness}, \binits{T.}},
\oauthor{\bsnm{Kanarek}, \binits{G.}},
\oauthor{\bsnm{Kendrew}, \binits{S.}},
\oauthor{\bsnm{Kern}, \binits{N.S.}},
\oauthor{\bsnm{Kerzendorf}, \binits{W.E.}},
\oauthor{\bsnm{Khvalko}, \binits{A.}},
\oauthor{\bsnm{King}, \binits{J.}},
\oauthor{\bsnm{Kirkby}, \binits{D.}},
\oauthor{\bsnm{Kulkarni}, \binits{A.M.}},
\oauthor{\bsnm{Kumar}, \binits{A.}},
\oauthor{\bsnm{Lee}, \binits{A.}},
\oauthor{\bsnm{Lenz}, \binits{D.}},
\oauthor{\bsnm{Littlefair}, \binits{S.P.}},
\oauthor{\bsnm{Ma}, \binits{Z.}},
\oauthor{\bsnm{Macleod}, \binits{D.M.}},
\oauthor{\bsnm{Mastropietro}, \binits{M.}},
\oauthor{\bsnm{McCully}, \binits{C.}},
\oauthor{\bsnm{Montagnac}, \binits{S.}},
\oauthor{\bsnm{Morris}, \binits{B.M.}},
\oauthor{\bsnm{Mueller}, \binits{M.}},
\oauthor{\bsnm{Mumford}, \binits{S.J.}},
\oauthor{\bsnm{Muna}, \binits{D.}},
\oauthor{\bsnm{Murphy}, \binits{N.A.}},
\oauthor{\bsnm{Nelson}, \binits{S.}},
\oauthor{\bsnm{Nguyen}, \binits{G.H.}},
\oauthor{\bsnm{Ninan}, \binits{J.P.}},
\oauthor{\bsnm{N{\"{o}}the}, \binits{M.}},
\oauthor{\bsnm{Ogaz}, \binits{S.}},
\oauthor{\bsnm{Oh}, \binits{S.}},
\oauthor{\bsnm{Parejko}, \binits{J.K.}},
\oauthor{\bsnm{Parley}, \binits{N.}},
\oauthor{\bsnm{Pascual}, \binits{S.}},
\oauthor{\bsnm{Patil}, \binits{R.}},
\oauthor{\bsnm{Patil}, \binits{A.A.}},
\oauthor{\bsnm{Plunkett}, \binits{A.L.}},
\oauthor{\bsnm{Prochaska}, \binits{J.X.}},
\oauthor{\bsnm{Rastogi}, \binits{T.}},
\oauthor{\bsnm{Janga}, \binits{V.R.}},
\oauthor{\bsnm{Sabater}, \binits{J.}},
\oauthor{\bsnm{Sakurikar}, \binits{P.}},
\oauthor{\bsnm{Seifert}, \binits{M.}},
\oauthor{\bsnm{Sherbert}, \binits{L.E.}},
\oauthor{\bsnm{Sherwood-Taylor}, \binits{H.}},
\oauthor{\bsnm{Shih}, \binits{A.Y.}},
\oauthor{\bsnm{Sick}, \binits{J.}},
\oauthor{\bsnm{Silbiger}, \binits{M.T.}},
\oauthor{\bsnm{Singanamalla}, \binits{S.}},
\oauthor{\bsnm{Singer}, \binits{L.P.}},
\oauthor{\bsnm{Sladen}, \binits{P.H.}},
\oauthor{\bsnm{Sooley}, \binits{K.A.}},
\oauthor{\bsnm{Sornarajah}, \binits{S.}},
\oauthor{\bsnm{Streicher}, \binits{O.}},
\oauthor{\bsnm{Teuben}, \binits{P.}},
\oauthor{\bsnm{Thomas}, \binits{S.W.}},
\oauthor{\bsnm{Tremblay}, \binits{G.R.}},
\oauthor{\bsnm{Turner}, \binits{J.E.H.}},
\oauthor{\bsnm{Terr{\'{o}}n}, \binits{V.}},
\oauthor{\bparticle{van} \bsnm{Kerkwijk}, \binits{M.H.}},
\oauthor{\bparticle{de~la} \bsnm{Vega}, \binits{A.}},
\oauthor{\bsnm{Watkins}, \binits{L.L.}},
\oauthor{\bsnm{Weaver}, \binits{B.A.}},
\oauthor{\bsnm{Whitmore}, \binits{J.B.}},
\oauthor{\bsnm{Woillez}, \binits{J.}},
\oauthor{\bsnm{Zabalza}, \binits{V.}}:
2018,
{The Astropy Project: Building an inclusive, open-science project and status of
  the v2.0 core package}.
\url{http://arxiv.org/abs/1801.02634}.
\end{botherref}
\endbibitem

\bibitem[\protect\citeauthoryear{Verscharen and Marsch}{2011}]{Verscharen2011}
\begin{barticle}
\bauthor{\bsnm{Verscharen}, \binits{D.}},
\bauthor{\bsnm{Marsch}, \binits{E.}}:
\byear{2011},
\batitle{{Apparent temperature anisotropies due to wave activity in the solar
  wind}}.
\bjtitle{Annales Geophysicae}
\bvolume{29}(\bissue{5}),
\bfpage{909}.
\doiurl{10.5194/angeo-29-909-2011}.
\burl{http://www.ann-geophys.net/29/909/2011/}.
\end{barticle}
\endbibitem

\end{thebibliography}

\IfFileExists{\jobname.bbl}{} {\typeout{}
\typeout{****************************************************}
\typeout{****************************************************}
\typeout{** Please run "bibtex \jobname" to obtain} \typeout{**
the bibliography and then re-run LaTeX} \typeout{** twice to fix
the references !}
\typeout{****************************************************}
\typeout{****************************************************}
\typeout{}}

\end{article}
\end{document}